\documentclass[preprint,aps,amsmath,pre,showpacs]{revtex4-1}
\usepackage[utf8]{inputenc} 
\usepackage{graphicx}
\usepackage{amsmath,amssymb,amsthm,amsfonts}
\usepackage{epsfig}
\usepackage{color}
\usepackage{xcolor}
\usepackage{ulem}

\definecolor{brandeisblue}{rgb}{0.0, 0.44, 1.0}

\newcommand\bluesout{\bgroup\markoverwith{\textcolor{brandeisblue}{\rule[0.5ex]{2pt}{1.5pt}}}\ULon}
\newcommand\redsout{\bgroup\markoverwith{\textcolor{red}{\rule[0.5ex]{2pt}{1.5pt}}}\ULon} 

\begin{document}

\title{Global synchronization of partially forced Kuramoto oscillators on Networks}

\author{Carolina A. Moreira and Marcus A.M. de Aguiar}
\email{corresponding author:aguiar@ifi.unicamp.br}

\affiliation{Instituto de Física F\'{\i}sica `Gleb Wataghin',
Universidade Estadual de Campinas, Unicamp\\ 13083-970, Campinas,
SP, Brazil}

\begin{abstract}

We study the synchronization of Kuramoto oscillators on networks where only a fraction of them is
subjected to a periodic external force. When all oscillators receive the external drive the system
always synchronize with the periodic force if its intensity is sufficiently large. Our goal is to
understand the conditions for global synchronization as a function of the fraction of nodes being
forced and how these conditions depend on network topology, strength of internal couplings and
intensity of external forcing. Numerical simulations show that the force required to synchronize the
network with the external drive increases as the inverse of the fraction of forced nodes. However,
for a given coupling strength, synchronization does not occur below a critical fraction, no matter
how large is the force.  Network topology and properties of the forced nodes also affect the
critical force for synchronization. We develop analytical calculations for the critical force for
synchronization as a function of the fraction of forced oscillators  and for the critical fraction
as a function of coupling strength. We  also describe the transition from synchronization with the
external drive to spontaneous synchronization.

\end{abstract}



\maketitle

\section{Introduction}
\label{intro}

Coupled biological oscillators are abundant in nature and often need to work in synchrony to
regulate physical activities, such as pacemaker cells in the heart \cite{Michaels704}, neurons in
regions of the brain  \cite{Pikovsky2003,gray1994,Mackay1997} and fireflies flashing collectively to
help females find suitable mates \cite{Moiseff181,Buck1976}. Artificial systems, such as
electrochemical oscillators \cite{Kiss2008} and coupled metronomes \cite{Pantaleone2002}, have also
been studied. There are evidences that synchronization also plays a key role in information
processing in areas on the cerebral cortex \cite{steriade1997,kelso2014}. Even the brain rest state
activity is characterized by local rhythmic synchrony that induces spatiotemporally organized
spontaneous activity at the level of the entire brain \cite{Deco2011}.

The model of coupled oscillators introduced by Kuramoto  \cite{Kuramoto1975} has become a paradigm
in the study of synchronization and has been extensively explored in the last years in connection
with biological systems, neural networks and the social sciences \cite{Rodrigues2016}. The model
consists of $N$ oscillators described by internal phases $\theta_i$ which rotate with natural
frequencies $\omega_i$ typically selected from a symmetric distribution. In the original model all
oscillators interact with each other according to the equations 
\begin{equation}
\dot{\theta_i} = \omega_i + \frac{\lambda}{N} \sum_{j=1}^{N} \sin (\theta_j - \theta_i),
\label{original}
\end{equation}
where $\lambda$ is the coupling strength and $i = 1, ..., N$.

Kuramoto analyzed the system in the limit where $N$ goes to infinity and showed that for small
values of the coupling parameter the oscillators continue to move as if they were independent.
However, as the coupling increases beyond a critical value, a finite fraction of oscillators
start to move together, a behavior termed spontaneous synchronization. This fraction increases
smoothly with the coupling, characterizing a second order phase transition in the limit of infinite
oscillators. For large enough coupling the whole system oscillates on the same frequency, as if it
were a single element.

Synchronization in many biological systems, however,  is not spontaneous, but frequently depends on
external stimuli. Information processing in the brain, for example, might be triggered by visual,
auditory or olfactory inputs \cite{Pikovsky2003}. Different patterns of synchronized neuronal firing
are observed in the mammalian visual cortex when subjected to stimuli \cite{gray1994}. In the
sensomotor cortex synchronized oscillations appear with amplitude and spatial patterns that depend
on the task being performed \cite{gray1994,Mackay1997}. Synchronization of brain regions not
directly related to the task in question can be associated to disorders like epilepsy, autism,
schizophrenia and Alzheimer \cite{Uhlhaas2006,Schmidt2015}. In the heart, cardiac synchronization is
induced by specialized cells in the sinoatrial node or by an artificial pacemaker that controls the
rhythmic contractions of the whole heart \cite{Reece2012}. The periodic electrical impulses
generated by pacemakers can be seen as an external periodic force that synchronizes the heart cells.
Another example of driven system is the daily light-dark cycle on the organisms \cite{Liu1997}. In
mammalians, cells specialized on the sleep control exhibit intrinsic oscillatory behavior whose
connectivity is still unknown \cite{Liu2011}.  The change in the light-dark cycle leads to a
response in the circadian cycle mediated by these cells, which synchronize via external stimulus.

A natural extension of the Kuramoto model, therefore, is to include the influence of an external
periodic force acting on the system \cite{Sakaguchi1988,Ott2008,Childs2008,Hindes2015}. In this work
we consider systems where the oscillators' interconnections form a network and where the  force acts
only on a fraction of the oscillators.  We are interested in the conditions for global
synchronization as a function of the fraction of nodes being forced and how it depends on network
topology. We show that the minimum force $F_{crit}$ needed for global synchronization scales as
$1/f$, where $f$ is the fraction of forced oscillators, and it is independent of the internal
coupling strength $\lambda$. However, in order to reach synchronization with fraction $f$ a minimum
internal strength is needed. The degree distributions of the network and the set of forced nodes
modify the $1/f$ behavior in heterogeneous networks.  We develop analytical approximations
	for $F_{crit}$  as a function of the fraction $f$ of forced oscillators  and for the minimum
	fraction $f_{crit}$ for which synchronization occurs as a function of $\lambda$. This paper is
organized as follows: in section \ref{model} we describe the partially forced Kuramoto model and
present the results of numerical simulations in section \ref{results}. In section \ref{approx} we
discuss the analytical calculations for $F_{crit}(f)$ and $f_{crit}(\lambda)$ that take into account
network topology and explain most of the simulations. We summarize our conclusions in section
\ref{conclusions}.

\section{The Forced Kuramoto Model on Networks}
\label{model}

Here we consider three modifications of the original Kuramoto model:  first, to include
the possibility that each oscillator interacts only with a subset of the other oscillators, the
system will be placed on a network whose topology defines the interactions \cite{Arenas2008}; second, we
include the action of an external periodic force  \cite{Sakaguchi1988,Ott2008,Childs2008} and; third, we
allow the external force to act only on a subset of the oscillators, representing the 'interface'
of the system that interacts with the 'outside' world, like the photo-receptor cells in the eye \cite{gray1994}. 

The system is described by the equations
\begin{equation}
\dot{\theta_i} = \omega_i + F \, \delta_{i,C} \sin(\sigma t - \theta_i) + 
\frac{\lambda}{k_i} \sum_{j=1}^{N} A_{ij} \sin (\theta_j - \theta_i) ,
\label{forced1}
\end{equation}
where $A_{ij}$ is the adjacency matrix defined by $A_{ij} = 1$ if oscillators $i$ and $j$ interact
and zero if they do not; $k_i$ is the degree of node $i$, namely $k_i = \sum_j A_{ij}$; $F$ and $\sigma$ are respectively the
amplitude and frequency of the external force; and $C$ is the subgroup of oscillators subjected to
the external force. We have also defined $\delta_{i,C} = 1$ if $i \in C$ and zero otherwise and 
we shall call $N_C$ the number of nodes in the set $C$. 

Following \cite{Childs2008} we get rid of the explicit time dependence by performing a change of
coordinates to analyse the dynamics in a referential frame corotating with the driving force:
\begin{equation}
\phi_i = \theta_i - \sigma t
\end{equation}
which leads to 
\begin{equation}
\dot{\phi_i} = \omega_i -\sigma - F \, \delta_{i,C} \sin \phi_i + \frac{\lambda}{k_i} \sum_{j=1}^{N} A_{ij} \sin (\phi_j - \phi_i),
\label{forced2}
\end{equation}

The behavior of the system depends now not only on the distribution of natural frequencies and
coupling intensity $\lambda$, but also on the network properties, on the intensity and frequency of
the external force and on the size and properties of the set $C$. The role of network characteristics in
the absence of external forcing has been extensively studied in terms of clustering \cite{Mcgraw2005,Mcgraw2007,Mcgraw2008},
assortativity \cite{Restrepo2014} and modularity \cite{Oh2005,Arenas2006,Arenas2007}.

The behavior of the system under an external force has also been considered for very large and fully
connected networks when the force acts on all nodes equally \cite{Childs2008}. The system exhibits a
rich behavior as a function of the intensity and frequency of the external force. In particular, it
has been shown that if the force intensity is larger than a critical value $F_{crit}$ the system may
fully synchronize with the external frequency. Among the questions we want to answer here are how
synchronization with the external force changes as we make $N_C < N$ and how does that depend on the
topology of the network and on the properties of the nodes in $C$. In particular we are interested
in studying how the critical intensity $F_{crit}$ of the external force increases as $N_C$ decreases
and if there is a minimum number of nodes that need to be excited by $F$ in order to trigger
synchronization. In the next section we show the results of numerical simulations considering three
network topologies (random, scale-free and fully connected). Analytical calculations that describe
these results will be presented next.

\section{Numerical Results}
\label{results}

In order to get insight into the general behavior of the system we present a set of simulations for
the following networks: (i) fully connected with $N=200$ nodes (FC200), (ii) fully connected with
$N=500$ (FC500); (iii) random Erdos-Renyi network with $N=200$ and average degree $\left\langle k 
\right\rangle = $  10.51 (ER200) and (iv) scale-free Barabasi-Albert network with $N=200$ (BA200)
computed starting with $m_0 = 11$ fully connected nodes and adding nodes with $m = 10$ links with 
preferential attachment, so that $\langle k \rangle =$  9.83. In all simulations we have considered a 
Gaussian distribution of natural frequencies $g(\omega)$ with null mean and standard deviation $a=1.0$
for the oscillators.

For the fully connected networks the critical value $\lambda_c$ for the onset of synchronization can
be estimated when $N \rightarrow \infty$ as $\lambda_c = 2a \sqrt{2/ \pi} \approx 1.6$. For finite
networks the calculation of $\lambda_c$ can be performed numerically (see, for example,
\cite{wang2015}) and we have checked that $\lambda_c=1.6$ is  a good approximation even for $N=100$
and for the other topologies we used. Full synchronization occurs only for larger values of
$\lambda$ and we define $\lambda_f$ as the value where $r=0.95$ and $\dot{\psi}< 10^{-2}$. Here we
are interested in scenarios where the system synchronizes spontaneously when $F=0$ and, therefore,
we set $\lambda$ above $ \lambda_f$ to assure full spontaneous synchronization. The coupling
strength $\lambda$ has an important role in the synchronization process, as we discuss below. For
each network type and fraction $f=N_C/N$ of nodes interacting with the external force we calculate
the minimum (critical) force necessary for synchronization with the external frequency.

In order to characterize the dynamics we use the usual order parameter
\begin{equation}
z = r e^{i \psi} = \frac{1}{N}\sum_{i=1}^N e^{i \phi_i},
\end{equation}
where $r=1$ indicates full synchronization and $\dot{\psi}$ the frequency of the collective
motion. We note that, since we are working on a rotating frame, synchronization with $\sigma$ will imply
$\dot{\psi} = 0$ whereas spontaneous synchronization $\dot{\psi} = - \sigma$.

\begin{figure}
	\includegraphics[clip=true,width=8cm]{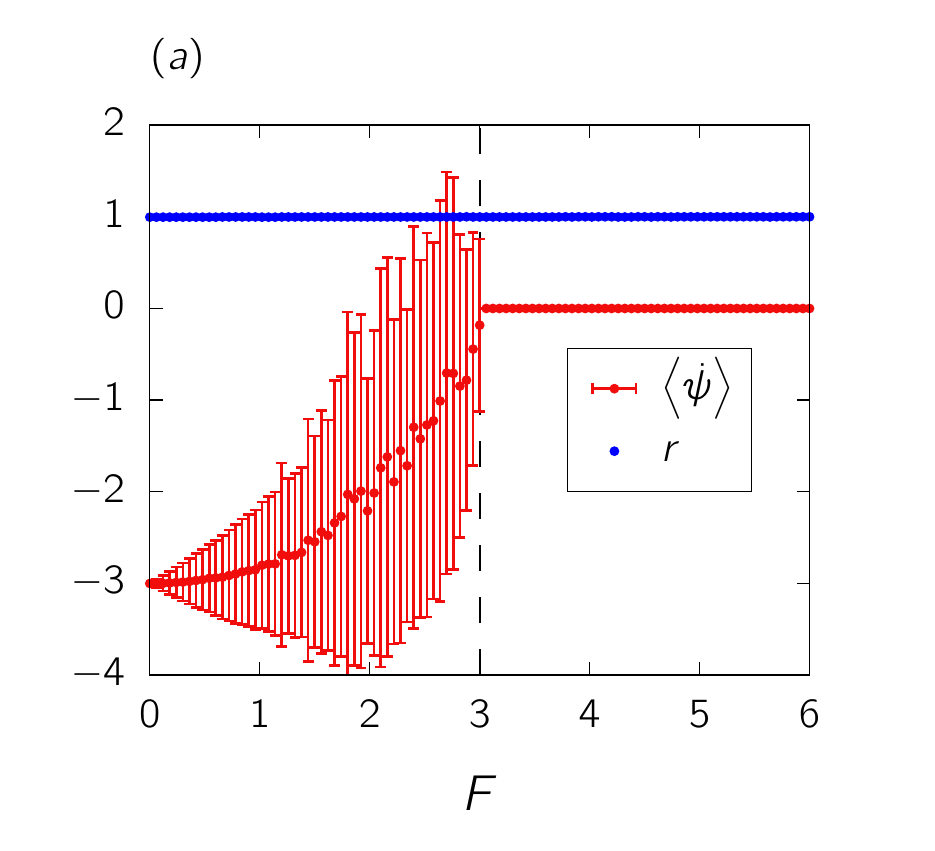}
	\includegraphics[clip=true,width=8cm]{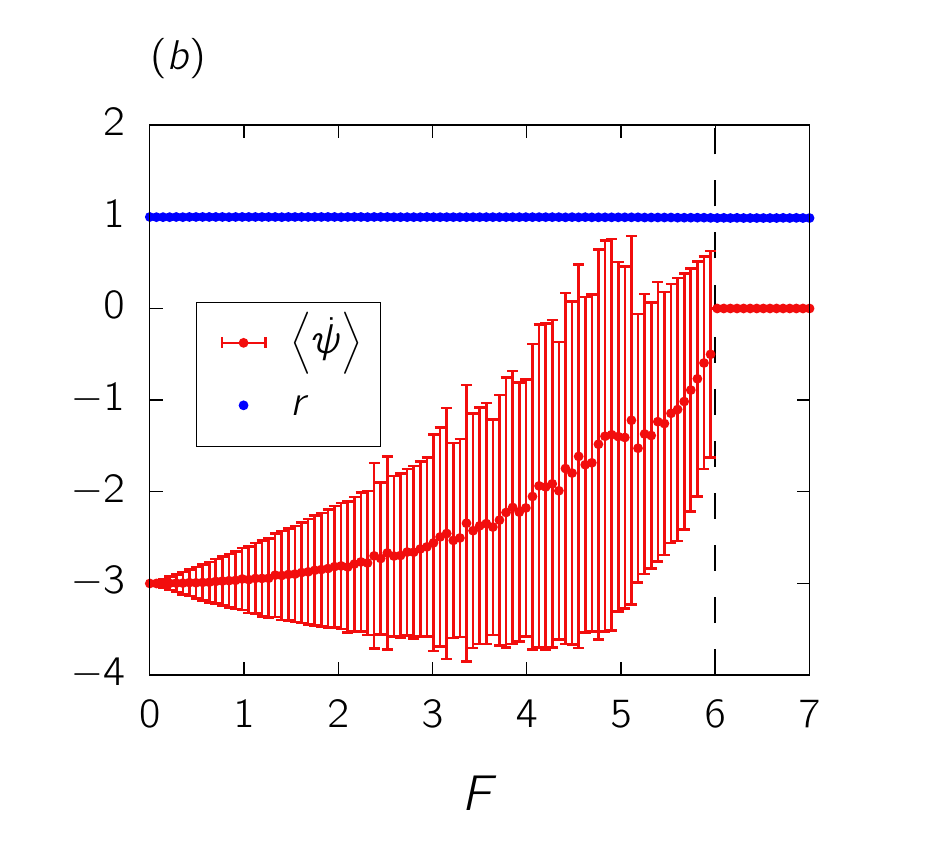} 
	\caption{(color online) Order parameter $r$ and $\dot{\psi}$ as a function of $F$ for a fully connected network
	with $N$ = 200, $\lambda$ = 20.0 and $\sigma$ = 3.0 for (a) $f = 1$ and (b) $f=0.5$. Red dots correspond to
	time averaged values calculated between $t=25$ to $t=50$. Error bars correspond to one standard deviation.
    The dashed lines indicate the critical force.} 
	\label{errorbars}
\end{figure}

Fig. \ref{errorbars} shows $r$ and $\dot{\psi}$ for FC200 as a function of $F$ for $\lambda = 20$
and $f=1$ and $f=0.5$. The system has been evolved up to $t=50$ starting with random phases, which
was enough to overcome the transient period. Because the system is finite and there are fluctuations
we computed time averages and standard deviations of $r$ and $\psi$  in the interval  from time $25$
to $50$.  The system remained fully synchronized for all values of $F$, first spontaneously ($F=0$)
and later with the external frequency for $F > 3$ ($f=1$) and for $F>6$ ($f=0.5$). For intermediate
values of the external force, $\dot{\psi}$ oscillates and the average and standard deviations are
shown. In this regime the oscillators move together ($r=1$) but change directions constantly due to
the competition between the couplings $\lambda$ and $F$. The critical force $F_{crit}$ was
numerically computed as the value of $F$ where $\dot{\psi} < 10^{-2}$ and $r > $ 0.95.

\begin{figure}
	\includegraphics[clip=true,width=8cm]{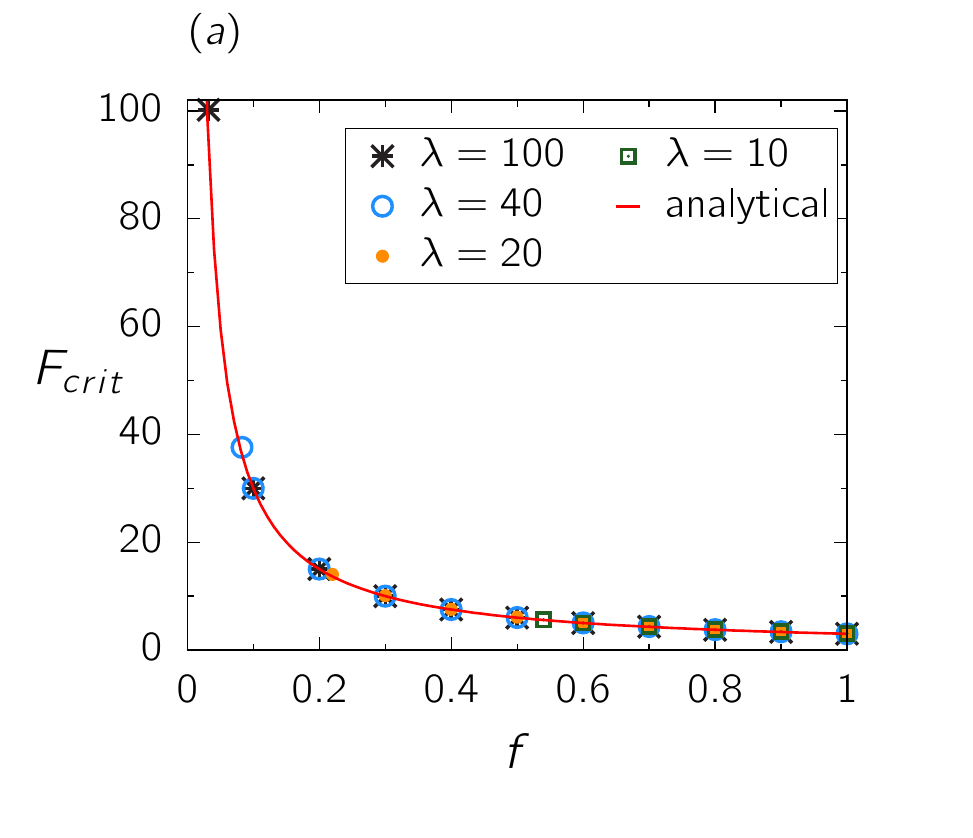}
	\includegraphics[clip=true,width=8cm]{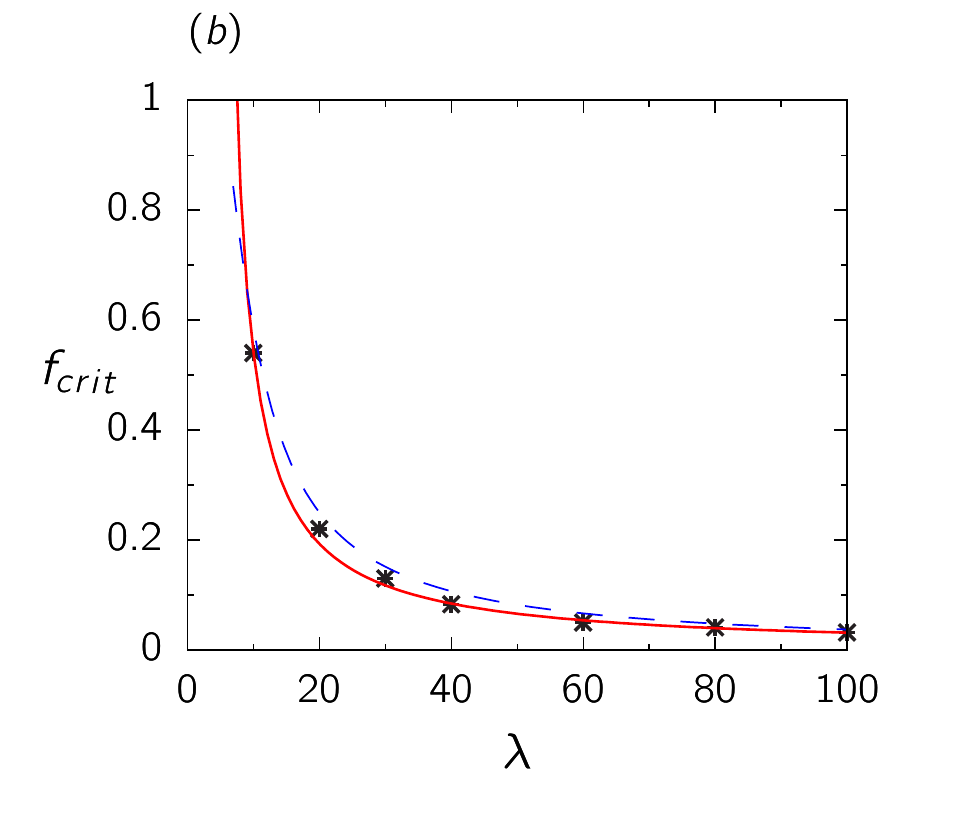}
	\caption{(color online) (a) Critical force $F_{crit}$ versus fraction $f$ of forced 
	nodes for the fully connected network FC200.  The continuous red curve shows the analytical 
	calculation and the symbols are the results of numerical simulations for different values
	of the coupling constant $\lambda$. The point with smallest $f$ for each $\lambda$ is 
	defined as $f_{crit}$. (b)  $f_{crit}(\lambda)$ from numerical simulations (stars)
	and according to  Eq. (\ref{fcrit0}) (red curve). The dashed (blue) line was obtained from 
	the parametric curve of Eq. (\ref{parametriclambda}).} 
		\label{fig2}
\end{figure}

Fig. \ref{fig2}(a) shows $F_{crit}$ as a function of the fraction $f$ of excited nodes for  FC200.
It also shows that for a fixed value of the internal coupling $\lambda$ synchronization can only be
achieved for $f$ larger than a critical value $ f_{crit}(\lambda)$. For example, for $\lambda=20$
(orange circles) synchronization is obtained only for $f > 0.22$. For $f < 0.22$ no synchronization
is achieved for $\lambda=20$, no matter how large is the external force. The value of $f_{crit}$ is
shown as the last point of the corresponding symbol on the plot.  Notice that the minimum value of
$F$ for synchronization does not itself depend on $\lambda$, since the same value is obtained as
long as $\lambda$ is large enough. Fig. \ref{fig2}(b) shows $f_{crit}$ as a function of $\lambda$.
We have performed the same analysis for FC500 and both curves $F_{crit}(f)$ and $f_{crit}(\lambda)$
were essentially identical to the ones obtained for FC200,  showing that these are independent of
network size.\\

\begin{figure}
	\includegraphics[clip=true,width=8cm]{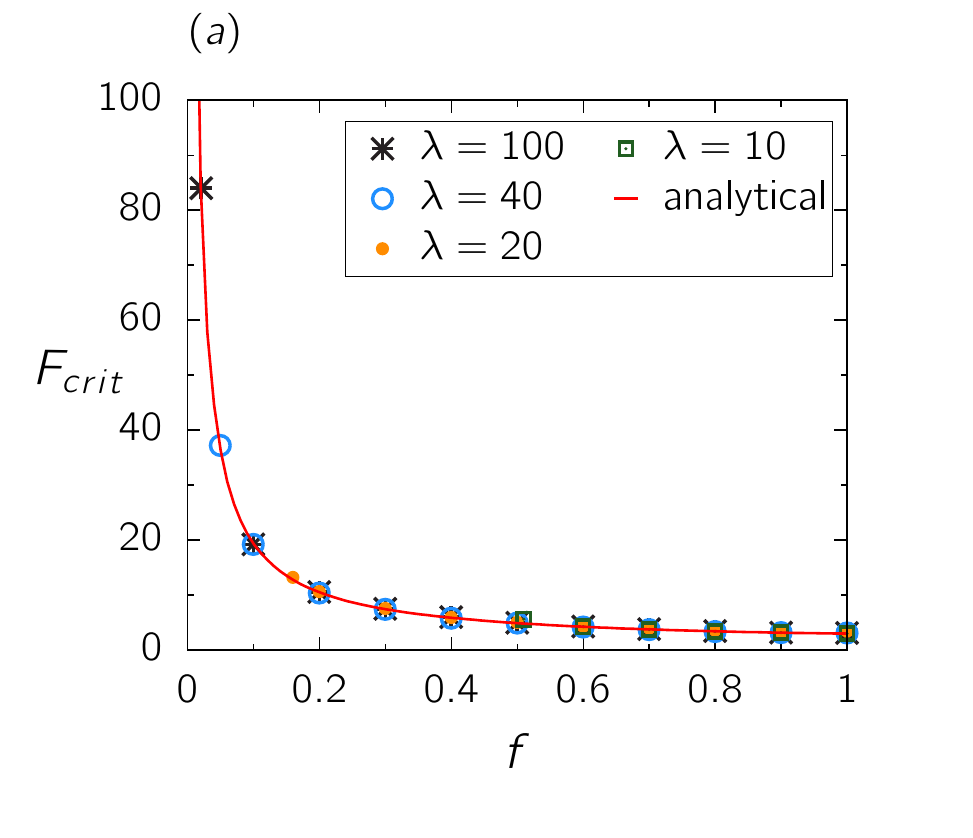}
	\includegraphics[clip=true,width=8cm]{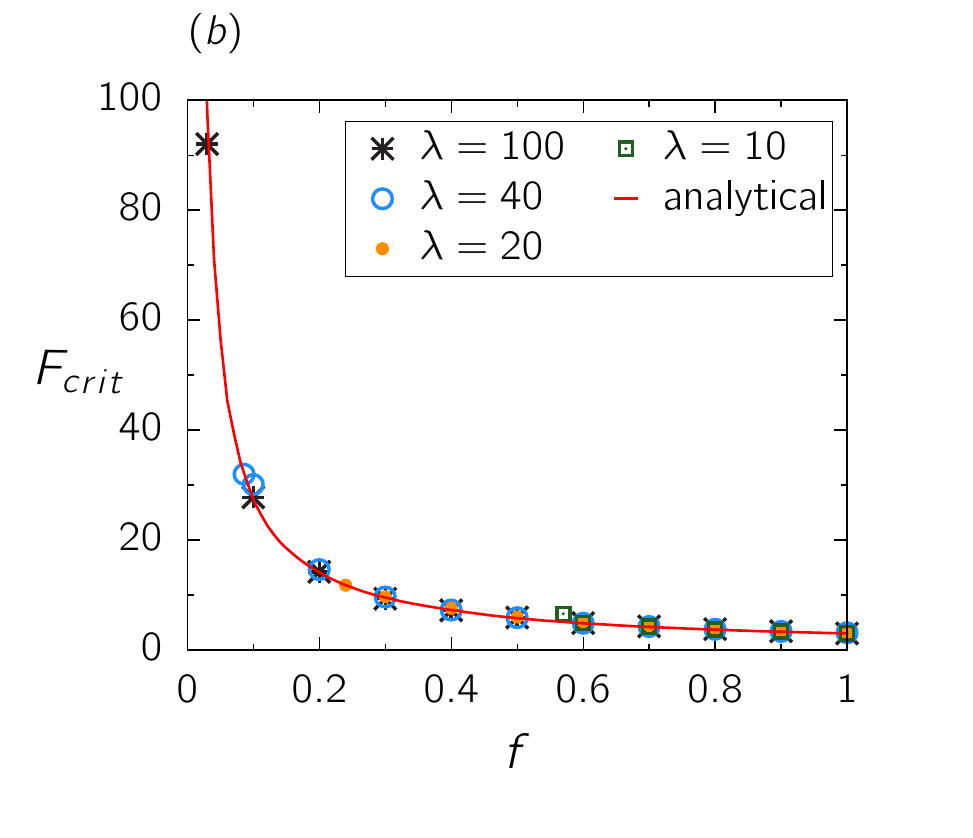}
	\caption{(color online) Critical force $F_{crit}$ versus fraction $f$ of forced nodes for the random network 
		ER200. The continuous red curve shows the analytical calculation and the symbols are the results of numerical simulations for different values of the coupling constant $\lambda$. The point with smallest $f$ for each $\lambda$ is defined as $f_{crit}$. Force is connected with nodes of (a)  highest degrees; (b) random. For the red line on panel (b) we have computed the average degree $\langle k \rangle_C$ of forced set  over 10 simulations to eliminate fluctuations.}
	\label{fig3}
\end{figure}

Fig. \ref{fig3} shows similar results for the ER200 random network. In this case the nodes have
different degrees and it matters which nodes are selected to interact with the external force. For the
results in panel (a) the nodes have been ordered from high to low degree and the $fN$ first (highly
connected) nodes have been selected to interact with the force. In panel (b) the nodes were chosen
at random.  The dependence of $f_{crit}$ on $\lambda$ is similar to the fully connected
case and different values of $\lambda$ are shown with different symbols.

\begin{figure}
	\includegraphics[clip=true,width=5.3cm]{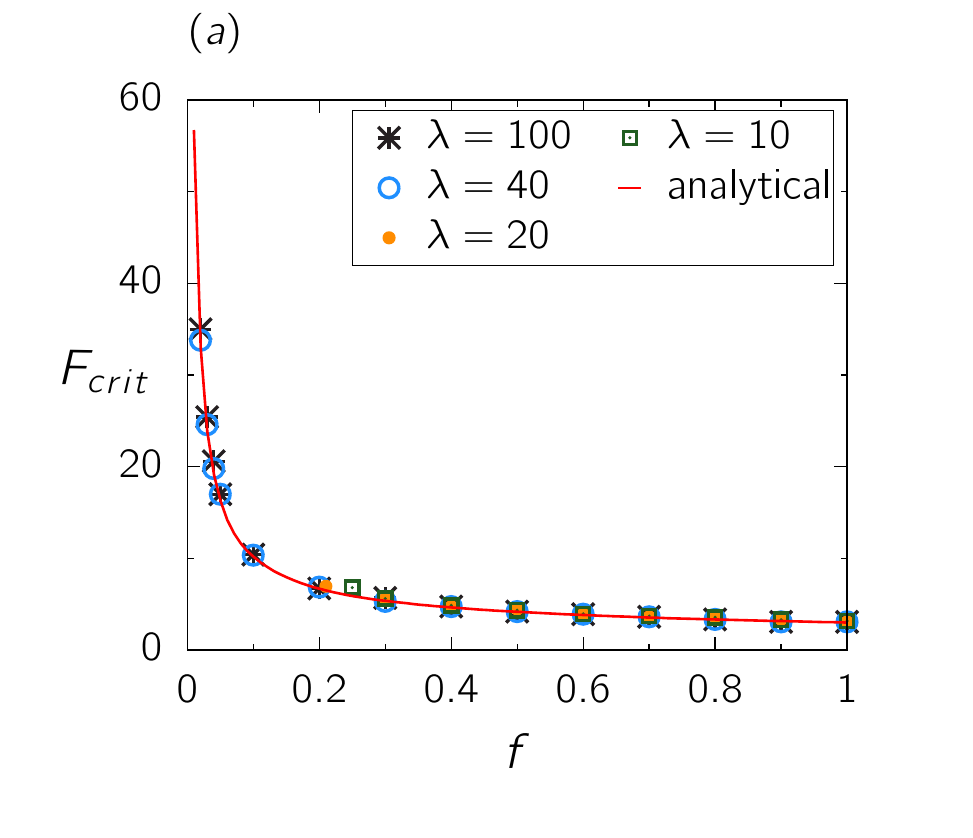}
	\includegraphics[clip=true,width=5.3cm]{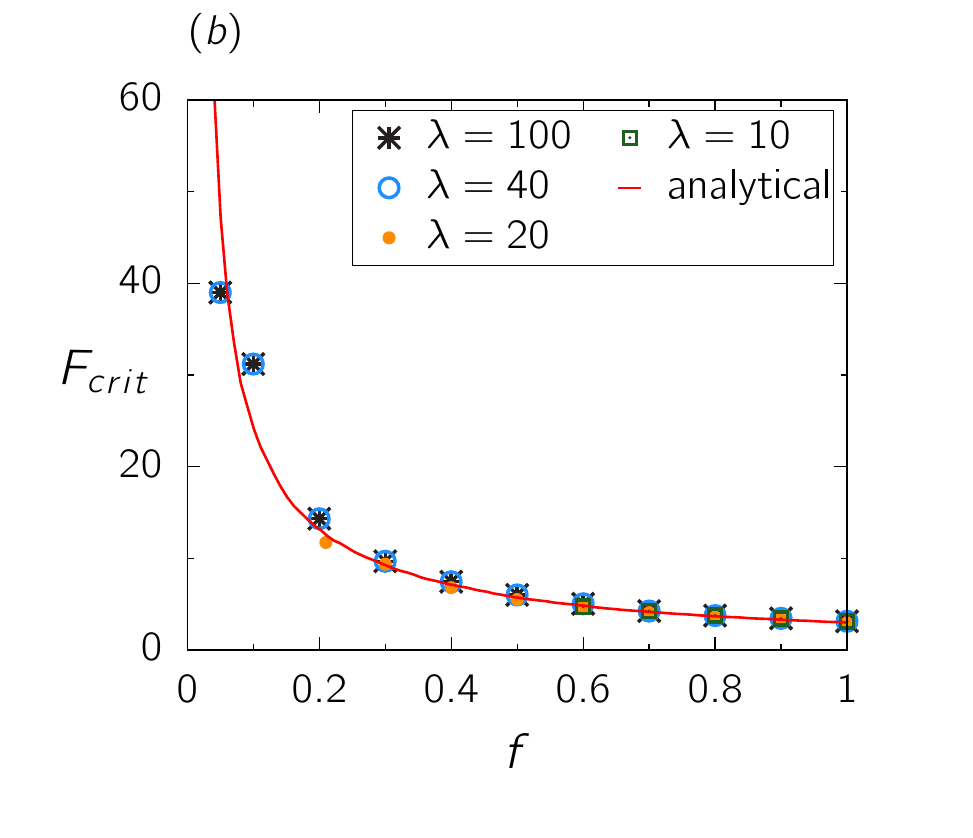}
	\includegraphics[clip=true,width=5.3cm]{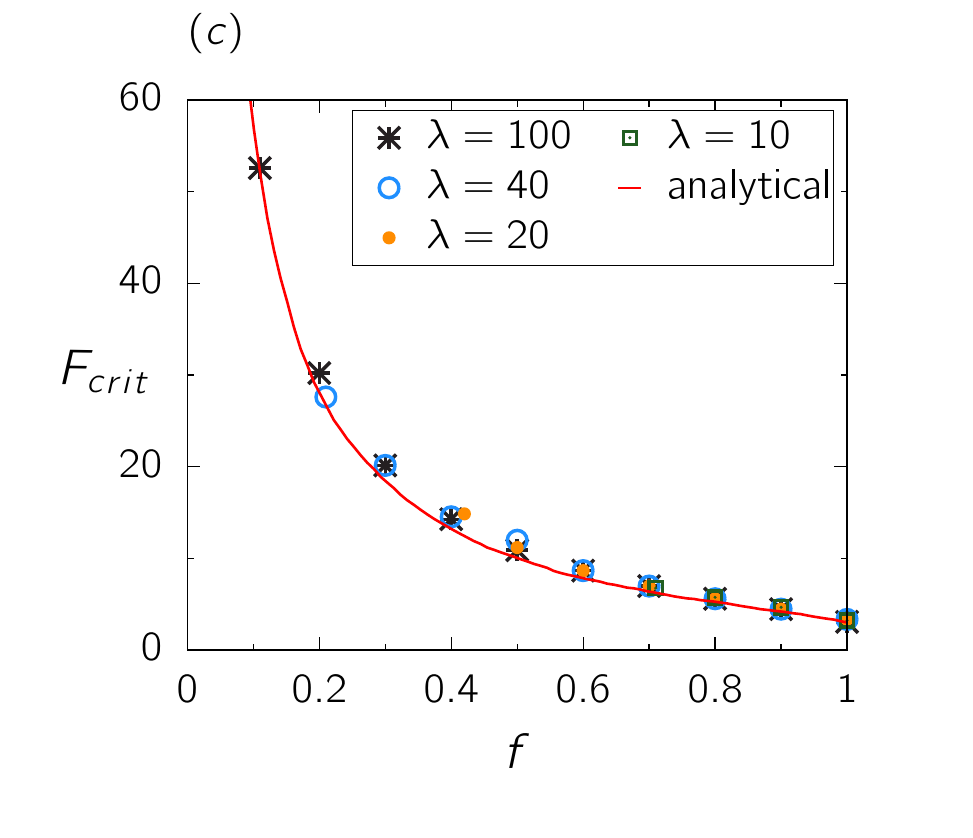}
	\caption{(color online) Critical force $F_{crit}$ versus fraction $f$ of forced nodes for the scale-free network 
		BA200. The continuous red curve shows the analytical calculation and the symbols are the results of numerical
		simulations for different values of the coupling constant $\lambda$. The point with smallest $f$ for each 
		$\lambda$ is defined as $f_{crit}$. Force is connected with nodes of (a)  highest degrees; (b) random and  
		(c) lowest degree. For the red line on panel (b) we have computed the average degree $\langle k \rangle_C$ of forced set  
		over 10 simulations to eliminate fluctuations.}
	\label{fig4}
\end{figure}

For the random network the differences between the two cases are not striking, since the
distribution of nodes is quite homogeneous. This is not the case for the BA200 network, as shown in
Fig. \ref{fig4}. When the  external source connects with nodes of highest degree, panel (a), the
critical force for synchronization is smaller  than when connected randomly, panel (b), or with
nodes of lowest degrees, panel (c), as expected. The analytical (red) curve for random connections
shows an average over 10 simulations using the same network but different random choices of nodes.

\section{Analytical results}
\label{approx}

The numerical simulations show that: (i) $F_{crit}$ depends of $f$; (ii) for heterogeneous networks
it depends on the properties of the set $C$; (iii) there is a critical fraction $f_{crit}$, that
depends on the network type, on $C$ and on $\lambda$, below which no synchronization is possible. In
this section we derive a theory for $F_{crit}(f)$ and an approximation for $f_{crit}(\lambda)$.

\subsection{Critical Force $F_{crit}$}

In order to derive an expression for $F_{crit}(f)$ we use the fact that nodes directly
affect all their neighbors and, therefore, their importance should be proportional to their degree.
We start by multiplying all terms of Eq.(\ref{forced2}) by $k_i/\langle k \rangle$, sum over $i$ and divide by
$N$ to obtain 
\begin{equation}
\frac{d \langle \phi \rangle_k}{dt} = \langle \omega \rangle_k -\sigma  - \bar{F}  \langle \sin \phi \rangle_{k,C}
\label{ap3}
\end{equation}
where
\begin{equation}
\langle \phi \rangle_k = \frac{1}{N}  \sum_{i=1}^N  \frac{k_i}{\langle k \rangle} \phi_i ,
\end{equation}
\begin{equation}
\langle \omega \rangle_k = \frac{1}{N}  \sum_{i=1}^N  \frac{k_i}{\langle k \rangle} \omega_i 
\end{equation}
and
\begin{equation}
\langle \sin \phi \rangle_{k,C} = \frac{1}{N_c}  \sum_{i \in N_c}  \frac{k_i}{\langle k \rangle} \sin \phi_i .
\label{sinforc}
\end{equation}
The  term proportional to $\lambda$,  containing the coupling between the oscillators, cancel out
exactly. When the oscillators synchronize with the external force Eq.(\ref{sinforc}) becomes
\begin{equation}
\langle \sin \phi \rangle_{k,C} = \sin \langle \phi \rangle \frac{\langle k \rangle_C}{\langle k \rangle}
\end{equation}
and we define
\begin{equation}
\bar{F_k} = f \frac{\langle k \rangle_C}{\langle k \rangle} F.
\label{app1b}
\end{equation}
Since $\langle \phi \rangle$ is constant in the synchronized state Eq.(\ref{ap3}) implies
\begin{equation}
\sin \langle \phi \rangle = \frac{\langle \omega \rangle_k -\sigma}{ \bar{F_k}} . 
\label{apc}
\end{equation}
Because the $\omega_i$ are randomly distributed with zero average, $\langle \omega \rangle_k$ is
generally small for large networks (although not zero in a single realization of the frequency
distribution). The critical force is now estimated as $\bar{F}_c =\sigma -\langle \omega \rangle_k $
and 
\begin{equation}
F_{crit} = \frac{\sigma - \langle \omega \rangle_k }{f} \, \frac{\langle k \rangle}{\langle k \rangle_C} 
        \approx \frac{\sigma}{f} \, \frac{\langle k \rangle}{\langle k \rangle_C} .
\label{force_crit}
\end{equation}
For regular networks, in particular,  where all nodes have the same degree, 
$\langle k \rangle = \langle k \rangle_C$, the critical force is reduced to
\begin{equation}
F_{crit} = \frac{\sigma}{f}.
\label{apd}
\end{equation}

Eq. (\ref{force_crit}) shows that when nodes with high degree are being forced, $\left\langle k
\right\rangle_C > \left\langle k \right\rangle$,  the critical force for synchronization is smaller
than the value obtained by equation (\ref{apd}), since the external force is directly transmitted to
a large number of neighbors.  On the other hand, if $\left\langle k \right\rangle_C < \left\langle k
\right\rangle$ (nodes with low degree are being forced) the critical force must be higher than that
estimated by (\ref{apd}), since these nodes have few neighbors. This agrees with the results shown
in Figs. \ref{fig2}-\ref{fig4}  where the continuous (red) line shows the approximation
Eq.(\ref{force_crit}).  For the scalefree network, in particular, when the force acts on nodes of
highest degree, Fig. \ref{fig4}(a), $F_{crit} \approx 5$ for $f = 0.4$, whereas $F_{crit} \approx 15$ for the
same value of $f$ when the force acts on the nodes with smallest degree Fig. \ref{fig4}(c).

\subsection{The critical fraction $f_{crit}(\lambda)$}

Eq.(\ref{ap3}) is exact and it might appear to be completely independent of $\lambda$. This,
however, is not true, since the dynamics of the angles $\phi$ are implicitly coupled by $\lambda$
and synchronization is only possible if $\lambda$ is large enough. As $f$ decreases the amplitude of
the external force needed for synchronization increases and if it gets too much larger than
$\lambda$ the oscillators start to move almost independently and synchronization is hindered.

An approximation for minimum value of $f$ that can lead to synchronization for a given $\lambda$ can
be obtained by setting the internal coupling strength per node to the intensity of the external
force, i.e., $\lambda \simeq F$. Along the curve $F=F_{crit}$ we may write  $\lambda \simeq \sigma
\langle k \rangle/ (f \langle k \rangle_C)$ (see Eq.(\ref{force_crit})) or, taking into account that
complete spontaneous synchronization starts at $\lambda_f$, we propose that $f_{crit}$ can be
estimated as 
\begin{equation}
f_{crit} (\lambda) = \frac{\sigma}{\lambda-\lambda_0} \; \frac{\langle k \rangle}{\langle k \rangle_C},
\label{fcrit0}
\end{equation}
where $\lambda_0$ is a fit parameter. For fully connected networks $\langle k \rangle = \langle k
\rangle_C$ and Eq. (\ref{fcrit0}) reduces to $f_{crit} (\lambda) = \sigma/(\lambda-\lambda_0)$. For
the red curve in Fig.\ref{fig2}(b) we obtained $\lambda_0=4.48 \pm 0.12$ which fits very well the
numerical results (black stars). Note that the value of $\lambda$ for $f=1$ is $\lambda_0+\sigma =
7.48$ for which we find $r=0.99$ for $F=0$ although $\dot{\psi}$ is still fluctuating. Full
spontaneous synchronization ($r>0.95$ and $\dot{\psi} < 10^{-2}$) only occurs for $\lambda=11.3$.

The heuristic approximation given by Eq.(\ref{fcrit0}) can be made more precise using the
bifurcation surfaces derived by Childs and Strogatz \cite{Childs2008} for the case where the
external force acts on all nodes. The derivation assumed a Lorentzian distribution for the
oscillator's natural frequencies, but is believed to be valid for a larger class of such
distributions. The full bifurcation diagram is divided into five regions but is dominated by only
two: one where the oscillators are locked to the same frequency as the external force and one with
mutual, spontaneous, synchronization. These two main regions are separated by saddle-node
bifurcations given in the $F$ versus $\sigma$ plane, for $\lambda$ fixed, by the parametric
equations
\begin{equation}
	F(\lambda,r) = \frac{\sqrt{2} r^2}{(1-r^2)^2} \sqrt{\lambda^2 (1-r^2)^3 + 2\lambda(r^4-4r^2+3) - 8}
\end{equation}
\begin{equation}
	\sigma(\lambda,r) = \frac{(1+r^2)^{3/2}}{2(1-r^2)^2} 
		\sqrt{\lambda (r^2-1)[\lambda (r^2-1)^2-4] - 4(r^2+1) }
\end{equation}
where $r$ varies from approximately $0.66$ to $1.0$. The resulting curve $F=F(\sigma)$ can be
approximated by the simple relation $F = \sigma$, as predicted by eq.({\ref{apd}}). This
approximation becomes exact as $\lambda$ goes to infinity, or when $r=1$ and $\dot{\psi}=0$.

Solving these equations for $F$ and $\lambda$ we obtain
\begin{equation}
\lambda(\sigma,r) = \frac{2}{(r^2-1)^2} + 2 \sqrt{\frac{r^4}{(r^2-1)^4}+\frac{\sigma^2(r^2-1)}{(r^2+1)^3}}
\label{parametriclambda}
\end{equation}
and $F(\sigma,r) = F(\lambda(\sigma,r) ,r)$. This new set of parametric equations results in the
critical curve $F=F(\lambda)$, for fixed $\sigma$. Finally, using eq.(\ref{apd}) $F=\sigma/f$ we can compute
$f=f(\lambda)$ with the parametric functions $(\lambda(\sigma,r),\sigma/F(\sigma,r))$).
This curve is shown as dashed (blue) line in Fig. \ref{fig2}(b)  
and differs from the heuristic approximation only for small values of $\lambda$.

\subsection{Transition from forced to mixed dynamics}

Synchronization with the external force is possible only if $F > F_{crit}$, estimated by
Eq.(\ref{force_crit}). If $F < F_{crit}$ the system's behavior is determined by the competition
between spontaneous and forced motion. The transition between these two regimes was studied in
detail in ref. \cite{Childs2008} for the case of infinitely many oscillators, all of which coupled
to the external drive. Here we present a simplified description of the transition using the
analytical approach developed above.

\begin{figure}
	\includegraphics[clip=true,width=8.1cm]{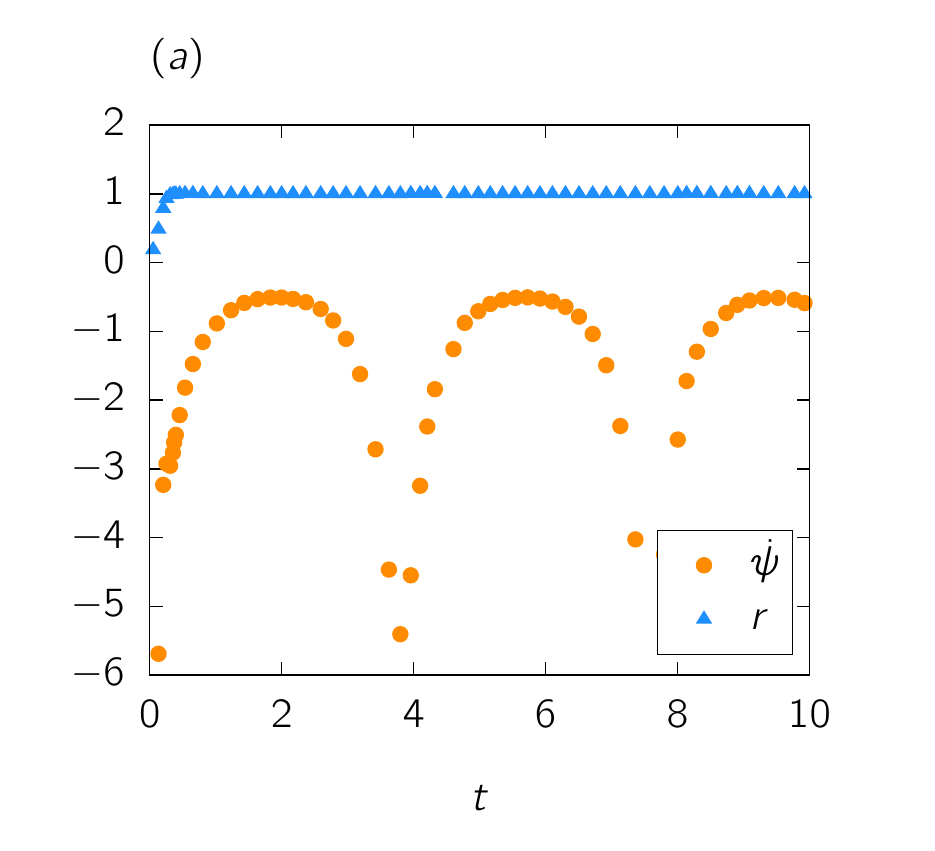}
	\includegraphics[clip=true,width=8.1cm]{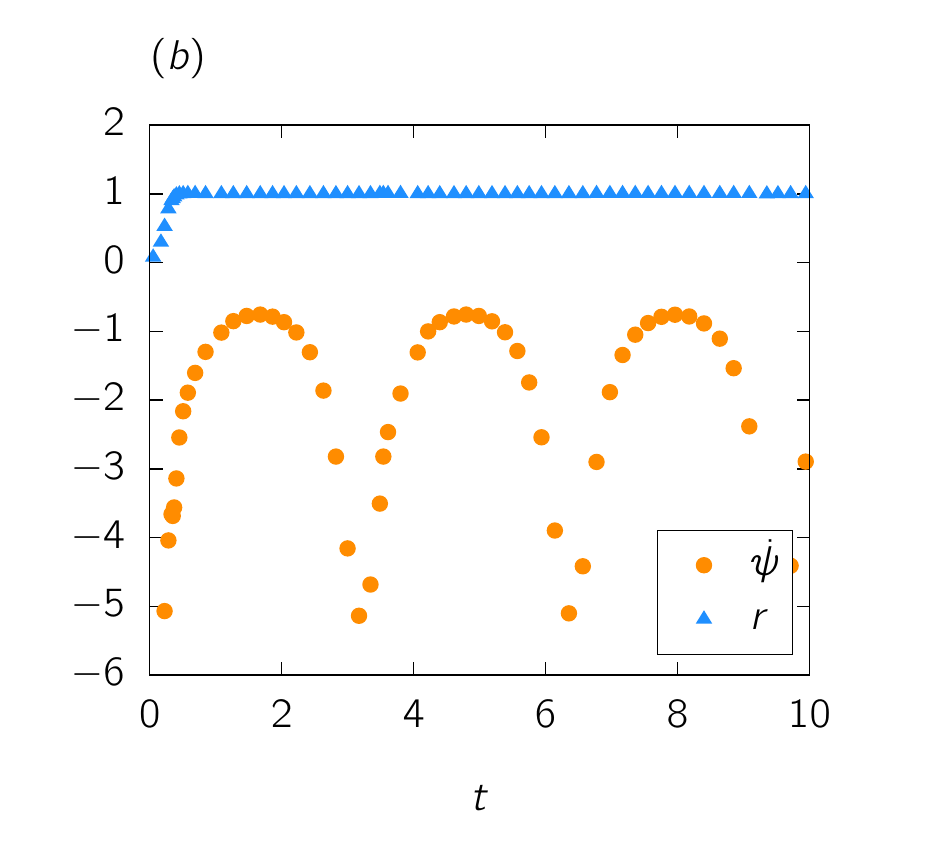}
	\includegraphics[clip=true,width=8.1cm]{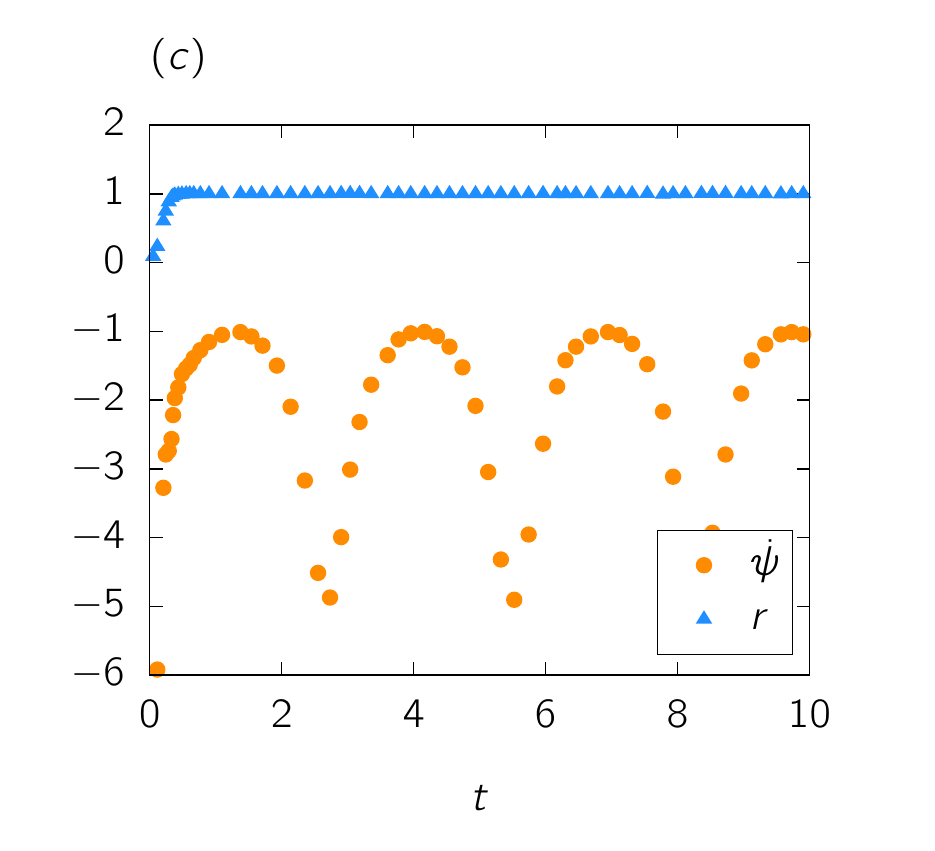}
	\includegraphics[clip=true,width=8.1cm]{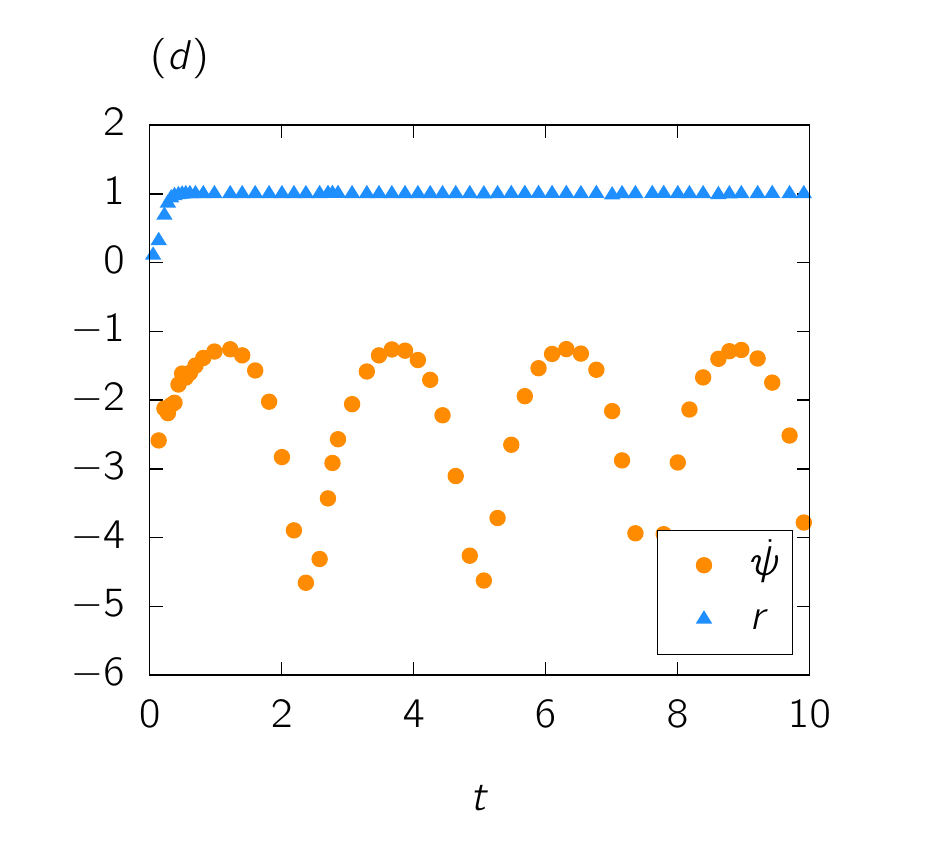}
	\caption{(color online) Frequency of oscillations for the fully connected network 
	with 200 nodes for $F=2.5$ fixed and fraction (a) $f=100 \%$; (b) $f=90 \%$; (c) $f=80 \%$ and 
	(d) $f=70 \%$. The points show $r$ (blue triangles) and $\dot{\psi}$ (orange circles). The 
	periods estimated from Eq.(\ref{period}) are (a) $\tau = 3.8$; (b) $\tau = 3.4$; (c) $\tau = 3.1$
	and (d) $\tau = 2.9$.}
	\label{fig5}
\end{figure}

Making the approximations $\langle \omega \rangle_k = 0$ and $\langle \sin \phi \rangle_{k,C} = \sin
\langle \phi \rangle$, Eq. (\ref{ap3}) simplifies to the Adler equation \cite{Adler1973}
\begin{equation}
\frac{d \phi }{dt} = -\sigma  - \bar{F}  \sin \phi 
\label{ap6}
\end{equation}
where we are omitting the average symbol and considering regular networks to simplify the notation. For general 
networks we only need to make $\bar{F} \rightarrow \bar{F}_k$. This equation, which has been used to 
model fireflies \cite{Ermentrout1984} among other systems \cite{Childs2008},  can be solved exactly to give
\begin{equation}
\sigma \tan{\phi/2} = \bar{F}  + \sqrt{\bar{F}^2 - \sigma^2} \tanh{\left[\frac{1}{2}\sqrt{\bar{F}^2 - \sigma^2}(t-t_0)\right]}
\end{equation}
for $\bar{F} > \sigma$. In this case $\phi$ converges to a constant value and the system stops (synchronizes with
$F$). For $\bar{F} < \sigma$, on the other hand, the solution is oscillatory,
\begin{equation}
\sigma \tan{\phi/2} = \bar{F}  - \sqrt{\sigma^2-\bar{F}^2} \tan{\left[\frac{1}{2}\sqrt{\sigma^2 - \bar{F}^2}(t-t_0)\right]}
\end{equation}
with period \cite{Jensen2002}
\begin{equation}
\tau = \frac{2\pi}{\sqrt{\sigma^2 - \bar{F}^2}}.
\label{period}
\end{equation}
Figure \ref{fig5} illustrates the frequency of oscillations for $F < F_{crit} = 3$ fixed and different number
of nodes that receive the external drive, showing $r$ and $\dot{\psi}$ as a function of $t$, for a fully
connected network. Although $r$ approaches 1 quickly (i.e., the system does synchronize), $\dot{\psi}$ 
oscillates with growing periods as the number of nodes on $C$ increases, remaining always negative. This means 
that $\psi$ decreases monotonically and the order parameter $z(t)$ oscillates, implying that a finite
fraction of the oscillators has synchronized spontaneously, due to their mutual interactions and not to the drive.
The approximation (\ref{period}) for the periods of oscillation matches very well the results of the
simulations.

\subsection{Time to equilibrium}

\begin{figure}
	\includegraphics[clip=true,width=11cm]{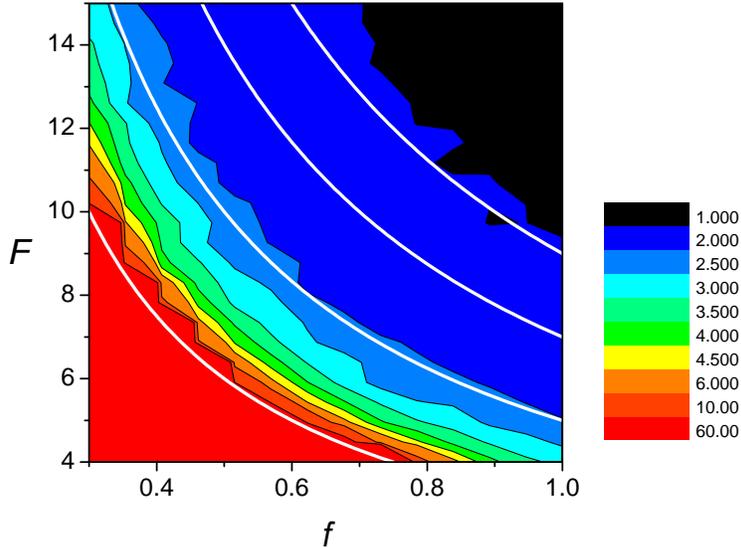}
	\caption{(color online) Contour plot of time to equilibration for different values of $F$ and $f$ and
	fixed $\lambda = 40$. Thick lines correspond to constant times according to the approximation
	$F=F_0/f$ for $F_0=3$, 5, 7 and 9.}
	\label{fig6}
\end{figure}

The time scale of dynamical processes also changes with the fraction of forced nodes. The time to
equilibrium should increase when $f$ decreases, but no simple relation seems to exist. When $F$
is large,  we can approximate Eq.(\ref{ap3}) by 
\begin{equation}
\frac{d \langle \phi \rangle_k}{dt} = - F f \frac{\langle k \rangle_C}{ \langle k \rangle}  \sin \langle \phi \rangle_k.
\label{forced4}
\end{equation}
Defining $t' = t F f \langle k \rangle_C/ \langle k \rangle$ this equation becomes identical to that
of a system where the force acts on all nodes. Therefore, within this crude approximation we expect
that: (i) for fixed $F$, the time to equilibration should scale as $\tau(f) = \tau_0 \, \langle k
\rangle/ [f \langle k \rangle_C]$, where $\tau_0$ is the equilibration time at $f=1$ and; (ii) along
the curve $F_{crit} (f) = F \, { \langle k \rangle} / [ f \langle k \rangle_C]$ the time to equilibration
remains constant, since the factors multiplying $F$ in Eq.(\ref{forced4}) cancel out. 
Fig.(\ref{fig6}) shows contour levels of numerically computed equilibration times in the $F \times f$ plane.
Thick black lines shows predicted curves of constant times, which indeed provide a somewhat poor
approximation to the computed values.

\section{Conclusions}\label{conclusions}

The Kuramoto model is perhaps the simplest dynamical system that allows the study of
synchronization. Here we considered the problem of periodically forced oscillators where the
external drive acts only on a fraction of them \cite{Childs2008}. The problem is inspired by
artificial heart pacemakers \cite{Reece2012} and information processing in the brain induced by an
external stimulus \cite{Pikovsky2003}. In both cases the stimulus is perceived by a subset of the
system (a heart chamber of photo-receptor cells in the eye) and propagates to other parts of the
network structure.

When the periodic drive acts on all oscillators, the system always synchronize with the forced
period if the force intensity is sufficiently large \cite{Childs2008}. Using numerical simulations
and analytical calculations we have shown that the force required to synchronize the entire set of
oscillators increases roughly as the inverse of the fraction of forced nodes. The degree
distribution of the complete network of interactions and of the set of forced nodes also affect the
critical force for synchronization. Forcing oscillators with large number of links facilitates
global synchronization in proportion to the average degree of the forced set to the total network.

We have also shown that below a critical fraction, that depends on $\lambda$, no synchronization
occurs, no matter how large the force. We believe this is an interesting result of this study that
might have consequences for adaptive systems relying on synchronization. The set of $N_C=fN$ nodes
that directly receives the external drive can be interpreted as the interface of a system where the
remaining $(1-f)N$ nodes are the 'processing unit', that needs to synchronize with the external
signal $F$ to perform a function. In this case it would be desirable to have $N_C$ as small as
possible to increase the processing power. However, synchronization with small $f$ requires large
couplings between the units, which can be costly. An example is neural network of {\it C. elegans}
where multiple links can connect the same two nodes and the cost of a connection is proportional to
the number of  such links (synaptic connections) that make it \cite{Latora2003}. In these cases it
is expected that a balance between interface size and network cost is attained, and natural systems
should evolve toward this condition. The final balance will, of course, depend on the cost. If the
cost is zero the system should evolve to the minimum possible interface size, given by $f=\sigma/F$
and $\lambda = \lambda_c + \sigma/f = \lambda_0+ F$ (for a fully connected network). If there is a
cost it might be advantageous to work with a smaller processing unit (and larger interface) that
requires smaller values of $\lambda$.

The theory developed here for $F_{crit}(f)$ considered only constant values of the coupling 
strength $\lambda$. In this case the term containing $\lambda$ in Eq.(\ref{forced2}) disappears from
the averaged Eq.(\ref{ap3}). This equation, however, remains valid for arbitrary symmetric couplings
$\lambda_{ij}=\lambda_{ji}$, as can be easily verified by inspection. For asymmetric couplings, 
$\lambda_{ij} \neq \lambda_{ji}$, this is not true and an extra term has to be included in
Eq.(\ref{ap3}). However, since this term is proportional to $\sin(\theta_j - \theta_i)$ it vanishes
when the system synchronizes and Eq.(\ref{force_crit}) still holds, being, therefore, a very robust
result.

As a final comment we note that here we have picked nodes for the set $C$ at random or based on
their degree. Another interesting choice would be to pick them according to their natural
frequencies $\omega_i$. For finite systems the oscillator with the largest frequency determines the
spontaneous synchronization of the system \cite{wang2015} and forcing the fastest nodes might also
result in interesting dynamics.\\

\noindent Acknowledgments: We thank Marlon R. Ferreira, David M. Schneider and Lucas D. Fernandes
for helpful discussions and suggestions. M.A.M.A.  acknowledges financial support from CNPq and
FAPESP. C.A.M. was supported by CNPq.

\clearpage 
\newpage

\end{document}